\documentclass[a4paper,fleqn,usenatbib]{mnras}
\usepackage{graphicx}
\usepackage{color}
\usepackage{rotating}
\usepackage{lscape}
\usepackage{morefloats}
\usepackage{amssymb}
\DeclareMathAlphabet{\mathpzc}{OT1}{pzc}{m}{it}

\def\lsim{\,\lower2truept\hbox{${<\atop\hbox{\raise4truept\hbox{$\sim$}}}$}\,}
\def\gsim{\,\lower2truept\hbox{${> \atop\hbox{\raise4truept\hbox{$\sim$}}}$}\,}

\title[ALMA calibrators]{ALMACAL IV: A catalogue of ALMA calibrator continuum observations}
\author[M. Bonato et al.]
{M. Bonato$^{1,2}$\thanks{bonato@ira.inaf.it},
E. Liuzzo$^{1}$,
A. Giannetti$^{1}$,
M. Massardi$^{1}$,
G. De Zotti$^{2}$,
S. Burkutean$^{1}$,
\newauthor
V. Galluzzi$^{3,4}$,
M. Negrello$^{5}$,
I. Baronchelli$^{6}$,
J. Brand$^{1}$,
M. A. Zwaan$^{7}$,
K. L. J. Rygl$^{1}$,
\newauthor
N. Marchili$^{8}$,
A. Klitsch$^{7,9}$
and I. Oteo$^{10,7}$\\
$^{1}$INAF$-$Istituto di Radioastronomia, and Italian ALMA Regional Centre, Via Gobetti 101, I-40129, Bologna, Italy\\
$^{2}$INAF$-$Osservatorio Astronomico di Padova, Vicolo Osservatorio 5, I-35122, Padova, Italy \\
$^{3}$INAF$-$Osservatorio Astronomico di Trieste, Via Tiepolo 11, I-34143, Trieste, Italy\\
$^{4}$Dipartimento di Fisica e Astronomia, Universit\`a di Bologna, via Gobetti 93/2, I-40129 Bologna, Italy\\
$^{5}$School of Physics and Astronomy, Cardiff University, The Parade, Cardiff CF24 3AA, UK\\
$^{6}$California Institute of Technology, Pasadena, CA\\
$^{7}$European Southern Observatory, Karl-Schwarzschild-Str. 2, D-85748 Garching, Germany\\
$^{8}$INAF$-$IAPS, Via Fosso del Cavaliere 100, I-00133 Roma, Italy\\
$^{9}$Centre for Extragalactic Astronomy, Department of Physics, Durham University, South Road, Durham DH1 3LE, UK\\
$^{10}$Institute for Astronomy, University of Edinburgh, Royal Observatory, Blackford Hill, Edinburgh EH9 3HJ, UK}

\pagerange{\pageref{firstpage}--\pageref{lastpage}} \pubyear{2013}

\def\LaTeX{L\kern-.36em\raise.3ex\hbox{a}\kern-.15em
    T\kern-.1667em\lower.7ex\hbox{E}\kern-.125emX}
\def\simlt{\mathrel{\rlap{\lower 3pt\hbox{$\sim$}}\raise 2.0pt\hbox{$<$}}}
\def\simgt{\mathrel{\rlap{\lower 3pt\hbox{$\sim$}}\raise 2.0pt\hbox{$>$}}}

\begin{document}

\label{firstpage}

\maketitle

\begin{abstract}
We present a catalogue of ALMA flux density measurements of 754 calibrators
observed between August 2012 and September 2017, for a total of 16,263
observations in different bands and epochs. The flux densities were measured
reprocessing the ALMA images generated in the framework of the ALMACAL
project, with a new code developed by the Italian node of the European ALMA
Regional Centre. A search in the online databases yielded redshift
measurements for 589 sources ($\sim$78 per cent of the total). Almost all
sources are flat-spectrum, based on their low-frequency spectral index, and
have properties consistent with being blazars of different types. To
illustrate the properties of the sample we show the redshift and flux density
distributions as well as the distributions of the number of observations of
individual sources and of time spans in the source frame for sources observed
in bands 3 (84$-$116\,GHz) and 6 (211$-$275\,GHz). As examples of the
scientific investigations allowed by the catalogue we briefly discuss the
variability properties of our sources in ALMA bands 3 and 6 and the frequency
spectra between the effective frequencies of these bands. We find that the
median variability index steadily increases with the source-frame time lag
increasing from 100 to 800 days, and that the frequency spectra of BL Lacs
are significantly flatter than those of flat-spectrum radio quasars. We also
show the global spectral energy distributions of our sources over 17 orders
of magnitude in frequency.

\end{abstract}


\section{Introduction}\label{sect:intro}

The Atacama Large Millimeter/submillimeter Array (ALMA) calibrators comprise
many hundreds of bright, compact radio sources, distributed
over about 85 per cent of the sky.

Every ALMA science project includes observations of calibrator sources (mostly bright quasars in the mm and sub-mm regime) to set the flux
density scale, to measure the bandpass response, and to calibrate amplitude and phase of the visibilities of the science targets \citep{Fomalont2014}.

Such observations represent a significant fraction (generally $\lesssim$30 per cent) of each execution block (EB). If the calibrator is a phase calibrator, it is observed many times during the same EB. If it is a bandpass or an amplitude calibrator, it is typically observed once per EB. Therefore each calibrator can be observed several times, on different dates, in different ALMA bands and array configurations, for one or multiple science projects.

The fields around ALMA calibrators of projects stored in the ALMA Science Archive have been exploited to carry out a novel, wide and deep (sub-)millimetre survey, ALMACAL \citep{Oteo2016}, and to investigate detected sources of special interest \citep{Oteo2017, Klitsch2017}.
The ALMACAL survey, in fact, takes advantage of the high sensitivity reached in the fields of ALMA calibrator observations to blindly extract a multi-band, multi-epoch survey of dusty star forming galaxies. Together with this primary goal, the same observations offer a unique opportunity to investigate spectral behaviour and variability of a large sample of bright extragalactic sources, the calibrators themselves, mostly AGNs across the whole (sub-)millimetric band.

In this paper we present a catalogue of observations of the ALMA calibrators collected, so far, for the ALMACAL project purposes.
Their multi-epoch, multi-frequency measurements over a poorly explored spectral
region constitute a rich database, well-suited for a variety of scientific
investigations, some of which will be more extensively detailed in future papers of our collaboration.

This paper is structured as follows. In Section~\ref{sect:data}, the catalogue
is introduced. An account of the source classification is given in
Section~\ref{sect:sed_classification}, where we also compare the
frequency spectra of BL Lacs and flat-spectrum radio quasars (FSRQs). In
Section~\ref{sect:properties}, we describe the main properties of the catalogue
and, as an example of its scientific exploitation, we briefly discuss the
variability properties of our sources in ALMA bands 3 and 6. In
Section~\ref{sect:sed_extension}, we present the global spectral energy
distributions (SEDs) of sources, built collecting data from online databases.
Finally, Section~\ref{sect:conclusions} contains a short summary of the paper.

\begin{figure}
\begin{center}
\includegraphics[trim=6.0cm 3.0cm 6.0cm 8.0cm,natwidth=610,natheight=642,width=0.25\textwidth]{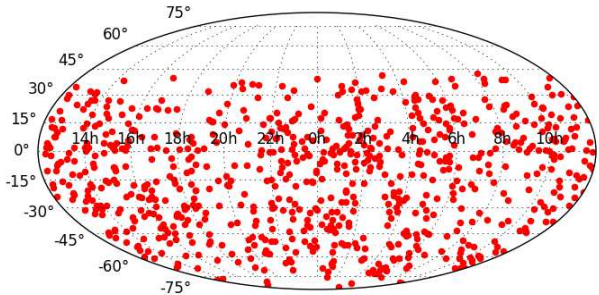}
\caption{Mollweide plot showing the spatial distribution of the ALMACAL calibrators considered in this paper.}
 \label{fig:Mollweide}
  \end{center}
\end{figure}

\begin{figure}
\begin{center}
\includegraphics[trim=2.2cm 1.0cm -0.2cm 14.5cm,natwidth=610,natheight=642,width=0.49\textwidth]{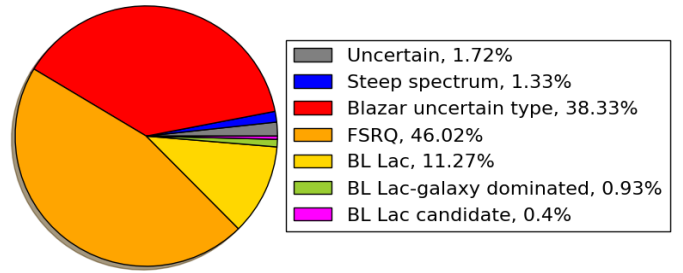}
\caption{Pie chart showing the fractions of our ALMACAL calibrators in the different classes of classification (see  Sect.\,\ref{sect:sed_classification}).}
 \label{fig:classification_pie}
  \end{center}
\end{figure}

\begin{table*}
\caption{Example of the catalogue content. The complete catalogue is available as supplementary material in the electronic version of the paper and on the website of the Italian ARC (\url{http://arc.ia2.inaf.it})}
\centering
\begin{tabular}{lccccccccc}
\hline
ALMA name & Class.$^{1}$ & z & RA [deg] & DEC [deg] & Flux density [Jy] & Error$^{2}$ [Jy] & band & $\nu$ [GHz] & Date of obs.$^{3}$\\
\hline
J1215-1731	 & 	4	 & 	0.669	 & 	183.9448	 & 	$-$17.5293	 & 	0.8602	 & 	0.043	 & 	3	 & 	95.4394	 & 	2013/03/16/07:19:31	 \\

 & & & & & 	0.3067	 & 	0.0153	 & 	7	 & 	340.686	 & 	2013/12/15/08:39:49	 \\

 & & & & & 	0.3017	 & 	0.0151	 & 	7	 & 	340.686	 & 	2013/12/15/10:03:31	 \\

 & & & & & 	0.7859	 & 	0.0393	 & 	3	 & 	95.4366	 & 	2014/04/03/07:14:35	 \\

 & & & & & 	0.7464	 & 	0.0373	 & 	3	 & 	95.4364	 & 	2014/04/05/03:54:26	 \\

 & & & & & 	0.7704	 & 	0.0385	 & 	3	 & 	95.4363	 & 	2014/04/05/05:12:34	 \\

 & & & & & 	0.4389	 & 	0.0219	 & 	6	 & 	225.342	 & 	2014/06/04/23:04:00	 \\

 & & & & & 	0.4141	 & 	0.0207	 & 	6	 & 	225.342	 & 	2014/06/05/01:11:37	 \\

 & & & & & 	0.7479	 & 	0.0374	 & 	3	 & 	112.496	 & 	2014/07/19/21:26:25	 \\

 & & & & & 	0.3421	 & 	0.0171	 & 	6	 & 	225.347	 & 	2014/08/16/17:37:15	 \\

 & & & & & 	0.3828	 & 	0.0191	 & 	6	 & 	225.347	 & 	2014/08/17/22:00:12	 \\

 & & & & & 	0.8924	 & 	0.0446	 & 	3	 & 	87.7719	 & 	2014/08/31/17:44:11	 \\

 & & & & & 	0.8265	 & 	0.0413	 & 	3	 & 	96.2087	 & 	2014/08/31/18:48:36	 \\

 & & & & & 	0.299	 & 	0.015	 & 	6	 & 	236.054	 & 	2016/03/03/04:19:24	 \\

 & & & & & 	0.3266	 & 	0.0163	 & 	6	 & 	234.085	 & 	2016/03/03/05:08:35	 \\

 & & & & & 	0.2093	 & 	0.0105	 & 	7	 & 	336.465	 & 	2016/09/15/14:43:43	 \\

 & & & & & 	0.1941	 & 	0.0097	 & 	7	 & 	336.465	 & 	2016/09/15/16:11:47	 \\

 & & & & & 	0.286	 & 	0.0143	 & 	6	 & 	226.385	 & 	2016/09/17/14:14:40	 \\

 & & & & & 	0.2708	 & 	0.0135	 & 	6	 & 	226.384	 & 	2016/09/17/15:28:17	 \\

 & & & & & 	0.276	 & 	0.0138	 & 	6	 & 	226.385	 & 	2016/09/18/13:53:52	 \\

 & & & & & 	0.255	 & 	0.0127	 & 	6	 & 	226.386	 & 	2016/09/22/17:58:47	 \\

 & & & & & 	0.3933	 & 	0.0197	 & 	4	 & 	138.666	 & 	2016/10/29/12:16:54	 \\

 & & & & & 	0.2432	 & 	0.0122	 & 	6	 & 	242.138	 & 	2016/11/11/11:15:06	 \\

 & & & & & 	0.264	 & 	0.0132	 & 	6	 & 	237.584	 & 	2016/11/19/12:35:00	 \\

 & & & & & 	0.4141	 & 	0.0207	 & 	4	 & 	138.672	 & 	2016/12/24/07:40:12	 \\

 & & & & & 	0.3785	 & 	0.0189	 & 	6	 & 	242.127	 & 	2017/03/18/02:30:48	 \\

 & & & & & 	0.2491	 & 	0.0125	 & 	7	 & 	348.498	 & 	2017/07/23/22:52:04	 \\

 & & & & & 	0.3571	 & 	0.0179	 & 	6	 & 	237.549	 & 	2017/08/08/20:29:36	 \\

 ... & ... & ... & ... & ... & ... & ... & ... & ... & ... \\
\hline
\multicolumn{10}{l}{\textit{Notes}. $^{1}$ Classification: 1=Flat-spectrum radio quasar (FSRQ); 2=BL Lac; 3=BL Lac-galaxy dominated; 4=Blazar uncertain type;}\\
\multicolumn{10}{l}{5=BL Lac candidate; 6=Steep spectrum; 7=Uncertain.}\\
\multicolumn{10}{l}{$^{2}$ The uncertainty is given by summing in quadrature the r.m.s. and a typical ALMA calibration error equal to 5 per cent of}\\
\multicolumn{10}{l}{the flux (see text).}\\
\multicolumn{10}{l}{$^{3}$ Observing time in the format [YYYY/MM/DD/hh:mm:ss], UTC time.}\\
\end{tabular}
\label{tab:fluxes}
\end{table*}

\begin{figure*}
\begin{center}
\includegraphics[trim=0.5cm 0.2cm 7.4cm 12.0cm,natwidth=610,natheight=642,width=0.49\textwidth]{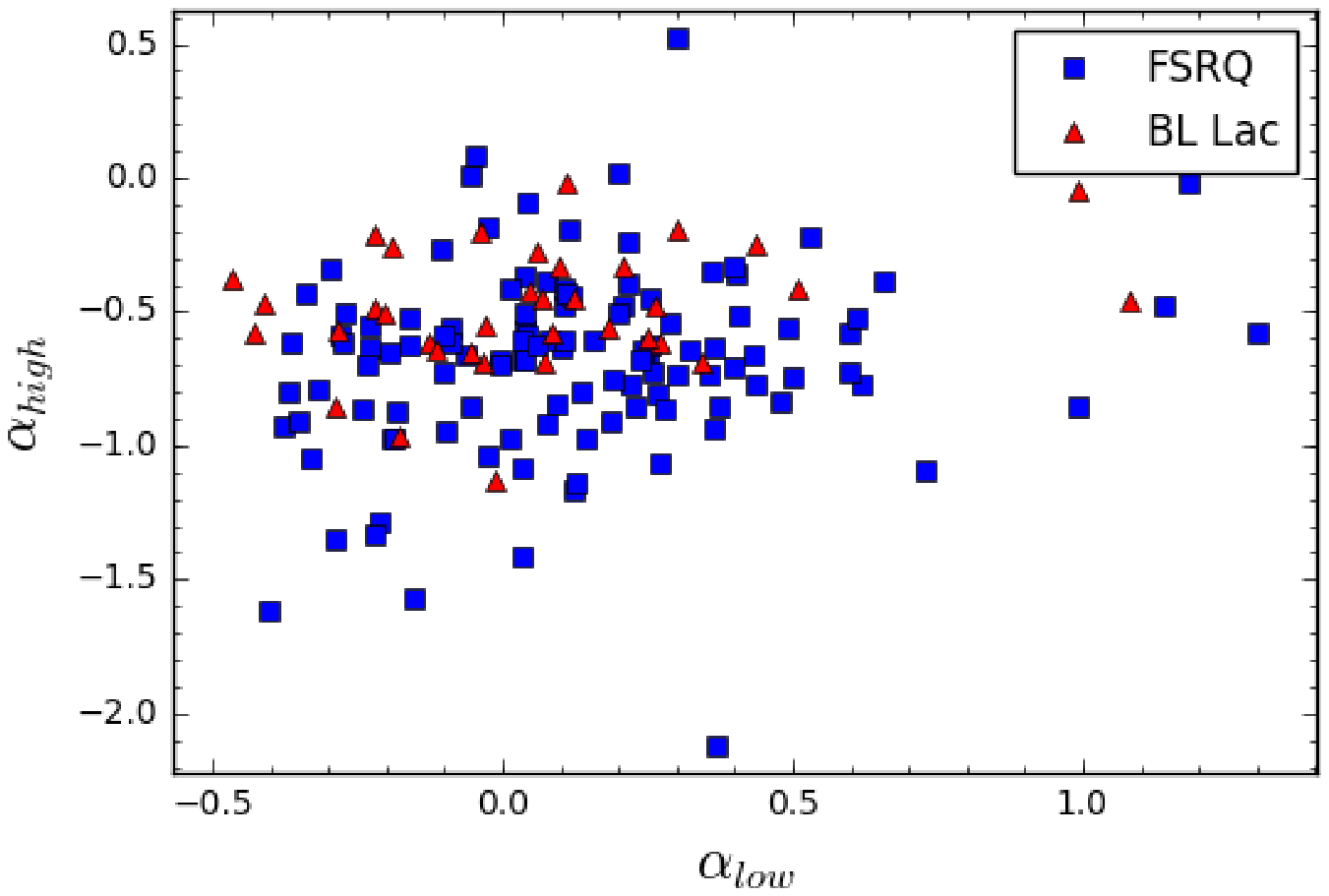}
\includegraphics[trim=0.5cm 0.2cm 7.4cm 12.0cm,natwidth=610,natheight=642,width=0.49\textwidth]{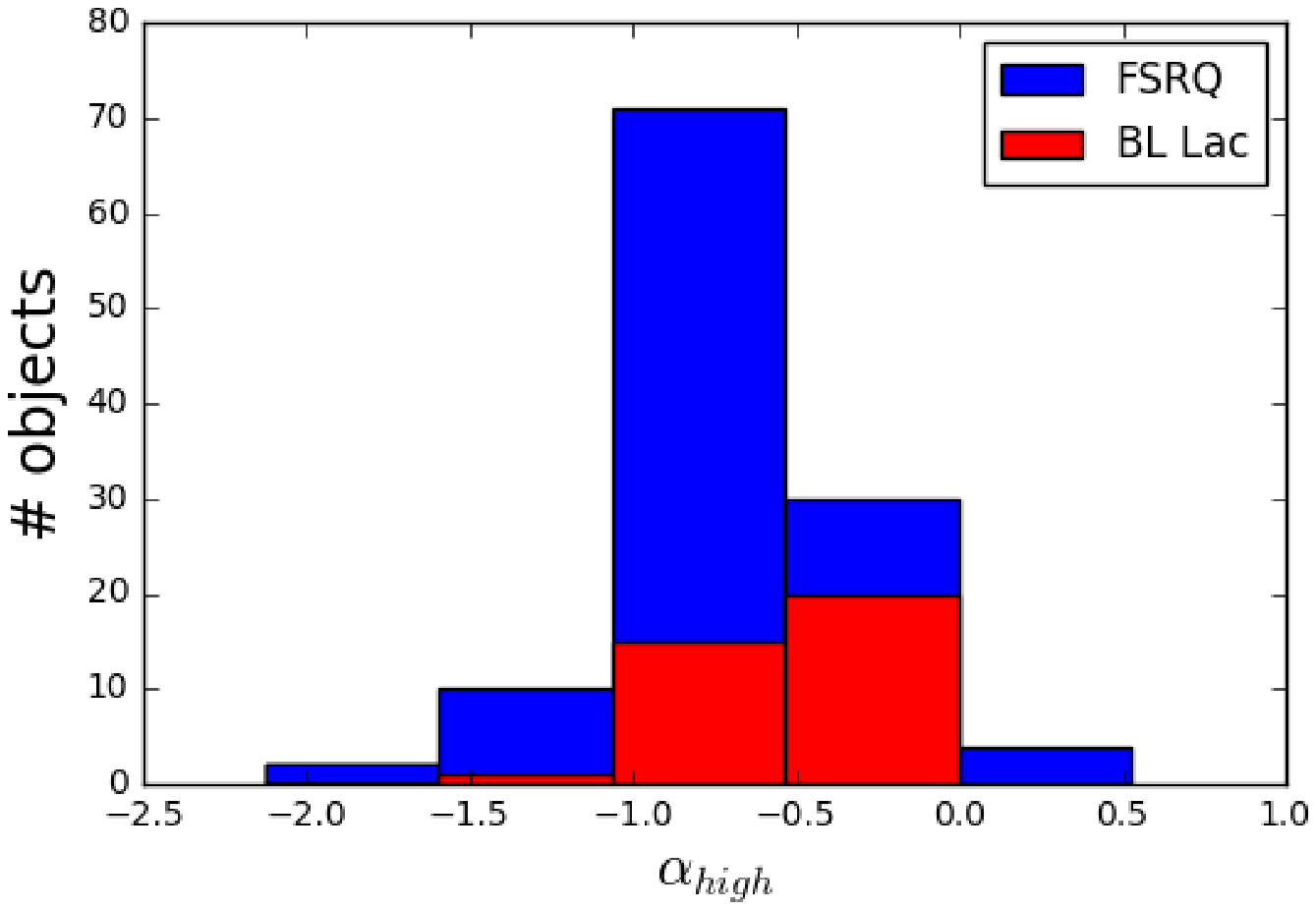}
\caption{On the left, high- versus low-frequency spectral indices of FSRQs and BL Lacs of our sample. $\alpha_{\rm low}$ is calculated between 1.4\,GHz (or 0.84\,GHz for sources outside the NVSS area) and 4.8\,GHz; $\alpha_{\rm high}$ is between the effective frequencies (listed in Table\,\ref{tab:fluxes}) in ALMA\,3 and in ALMA\,6 bands. The right panel shows the distribution of $\alpha_{\rm high}$ for the two populations.}
 \label{fig:color_color}
  \end{center}
\end{figure*}

\section{The sample }\label{sect:data}

The catalogue consists of continuum measurements of calibrators, obtained
during  the majority of the ALMA science observations between August 2012 and
September 2017. In total, we collected 16,263 observations\footnote{From the
initial sample, we removed $\sim$1.7 per cent of the images that showed
anomalies.} of 754 calibrators. Being a collection of data from a
heterogeneous sample of science projects, our observations vary within a wide
spectrum of different frequency setups, array configurations and integration
times. 

The details of calibration and imaging for the ALMACAL data are described by
\citet{Oteo2016}. Here we summarise a few pieces of information, useful for a comprehensive description of the presented catalogue.
For our purposes we considered all the ALMA projects in the epochs 2012-2017 and, in them, the extragalactic calibrators at any observing band.
The full data deliveries available in the ALMA archive were retrieved for datasets for which the proprietary period had expired, while only the calibrator data was considered for the remaining projects, after an official request through an ALMA Helpdesk ticket.
Calibration scripts produced during the ALMA Quality Assessment procedure and distributed through the archive were run to generate the calibration tables that were applied to all the calibrators (in some cases differently with respect to what usually done for the archived data, for which, expecially in the first observing cycles, tables were applied only to science targets and phase calibrators).
Data were self-calibrated taking advantage of the presence of the calibrator in the phase center and images are produced with the calibrator present and subtracted (in the visibility domain). The latter are used in the ALMACAL collaboration to investigate the background looking for dusty galaxies. The former are used in the present paper to investigate the calibrator population properties.

For the calibrators, the flux densities were uniformly measured from the
ALMA images\footnote{We derived the flux densities through an image analysis,
instead of simply using model fit values, because the former approach provides
robust measurements for both resolved and non-resolved sources, while model fit
flux densities are reliable for non-resolved observations only} using a new
code developed by the Italian node of the European ALMA Regional Centre (ARC).
This software is part of a suite of tools aimed at easing the ALMA Science
Archive mining: the ALMA Keyword Filler tool package (AKF; \citealt{Liuzzo18})
and the Keywords of Astronomical FITS-images Explorer (KAFE;
\citealt{Burkutean18}). The AKF codes are particularly useful to compare image
products or to identify the images to be selected for several scientific
purposes. KAFE is a web-based FITS image post-processing analysis tool. It
exploits AKF and complements selected FITS files with metadata based on a
uniform image analysis approach while also offering advanced image diagnostic
plots. KAFE's applicability to multi-instrument images in the radio to sub-mm
wavelength domain makes it ideal for data sample studies requiring uniform data
diagnostic criteria.

After the estimation of the r.m.s. ($\sigma$) in an image, the code masks
the pixels with a flux density below $5\,\sigma$ and obtains the source flux
density by integrating over the remaining pixels. This is enough to cope both with isolated point-like and extended sources, definition which strongly depends on the observing strategy and phase decoherence and might vary for our targets from one observation to the other.

The number of observations in the different ALMA bands are: 5100 in band 3 (84$-$116\,GHz), 639 in band 4 (125$-$163\,GHz),
6319 in band 6 (211$-$275\,GHz), 3584 in band 7 (275$-$373\,GHz), 393 in band 8
(385$-$500\,GHz), 220 in band 9 (602$-$720\,GHz) and 8 in band 10
(787$-$950\,GHz).

The ALMA measurements of the 754 calibrators are included as supplementary
material in the electronic version of the paper and on the website of the
Italian ARC (\url{http://arc.ia2.inaf.it}). The catalogue gives the ALMA name,
the source classification, its redshift (if available), the equatorial
coordinates (J2000), the flux density measured in each observation with its
error, the effective observing frequency, and the date and UTC time of the
observation. The error is essentially given by the uncertainty in the flux
density calibration (errors due to instrumental noise are typically
smaller by more than two orders of magnitudes); we adopt a calibration
uncertainty of 5 per cent (E. Fomalont, private communication)\footnote{The
debate about the precise value of the calibration uncertainty is still open in
the ALMA community. Our results about the differences in flux density of the different calibrators
for short time spans (see Sect.\,\ref{sect:properties}) support the adopted 5
per cent level.}. An example of the content of the catalogue is given in
Table\,\ref{tab:fluxes}. The coordinates are the average between the
positions measured in the different ALMA observations.

We have recovered the redshifts of 589 sources ($\sim$78 per cent of the
total), using the
Astroquery\footnote{\url{https://astroquery.readthedocs.io/en/latest/}}
affiliated package of astropy\footnote{\url{http://www.astropy.org/}} on the
NASA/IPAC Extragalactic database\footnote{\url{https://ned.ipac.caltech.edu/}}
(NED), VizieR\footnote{\url{http://vizier.u-strasbg.fr/viz-bin/VizieR}} and
SIMBAD\footnote{\url{http://simbad.u-strasbg.fr/simbad/}} databases. Redshifts
for 256 calibrators were provided by \citet{Mahony2011}. Whenever multiple
redshifts of the same source were found, we give the median value.

In Fig.\,\ref{fig:Mollweide}, we show the Mollweide projection of the positions
of the ALMA calibrators, obtained through KAFE.

\begin{figure}
\begin{center}
\includegraphics[trim=0.0cm 0.5cm 6.0cm 13.0cm,natwidth=610,natheight=642,width=0.49\textwidth]{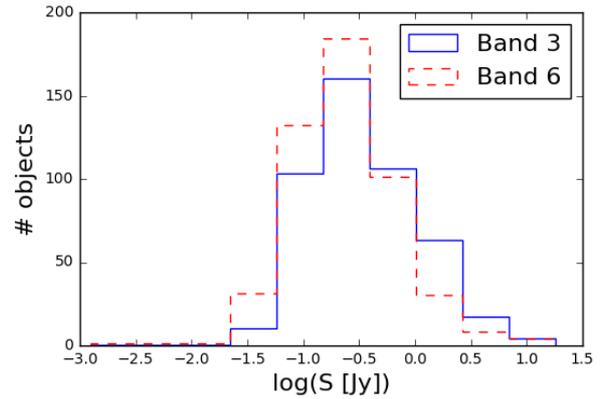}
\caption{Flux density distributions of sources detected in ALMA band 3 (solid blue line) and band 6 (dashed red line).}
 \label{fig:F_distribution}
  \end{center}
\end{figure}

\begin{figure}
\begin{center}
\includegraphics[trim=0.0cm 0.5cm 6.0cm 13.0cm,natwidth=610,natheight=642,width=0.49\textwidth]{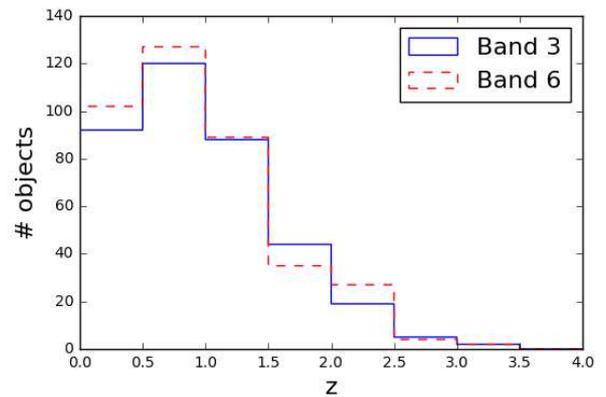}
\caption{Redshift distributions of the sources detected in ALMA band 3 (solid blue line) and band 6 (dashed red line).}
 \label{fig:z_distribution}
  \end{center}
\end{figure}

\begin{figure}
\begin{center}
\includegraphics[trim=0.0cm 0.5cm 6.0cm 13.0cm,natwidth=610,natheight=642,width=0.49\textwidth]{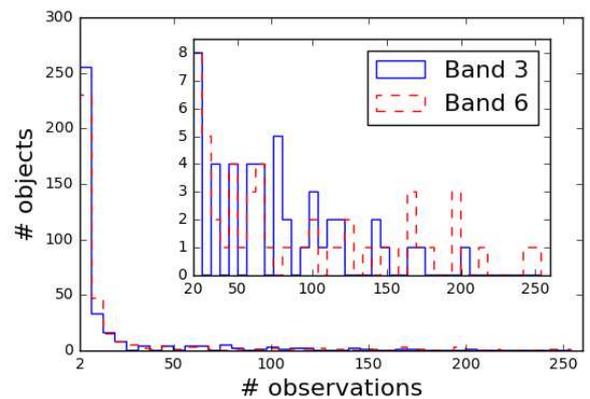}
\caption{Distributions of the number of observations per source in ALMA band 3 (solid blue line) and in band 6
(dashed red line); $\hbox{bin\ width}=6$. We only considered sources with $\geq$2 observations. In the zoomed-up inset plot, we show the portion of $\geq$20 observations only.}
 \label{fig:N_distribution}
  \end{center}
\end{figure}

\begin{figure}
\begin{center}
\includegraphics[trim=0.0cm 0.5cm 6.0cm 13.0cm,natwidth=610,natheight=642,width=0.49\textwidth]{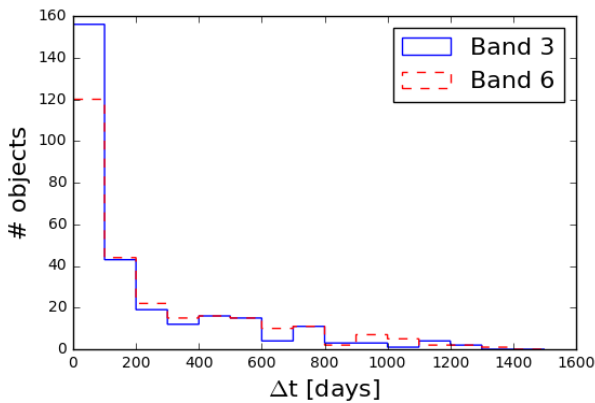}
\caption{Distributions of the rest-frame time spans of source observations (i.e.
$\Delta t=[t_{\rm last\,observation}-t_{\rm first\,observation}]/[1+z]$) in ALMA band 3
(solid blue line) and in band 6 (dashed red line). We only considered sources having redshift measurements and $\geq$2 observations.}
 \label{fig:time_distribution}
  \end{center}
\end{figure}

\begin{figure*}
\begin{center}
\includegraphics[trim=1.0cm 0.8cm 5.8cm 12.0cm,natwidth=610,natheight=642,width=0.32\textwidth]{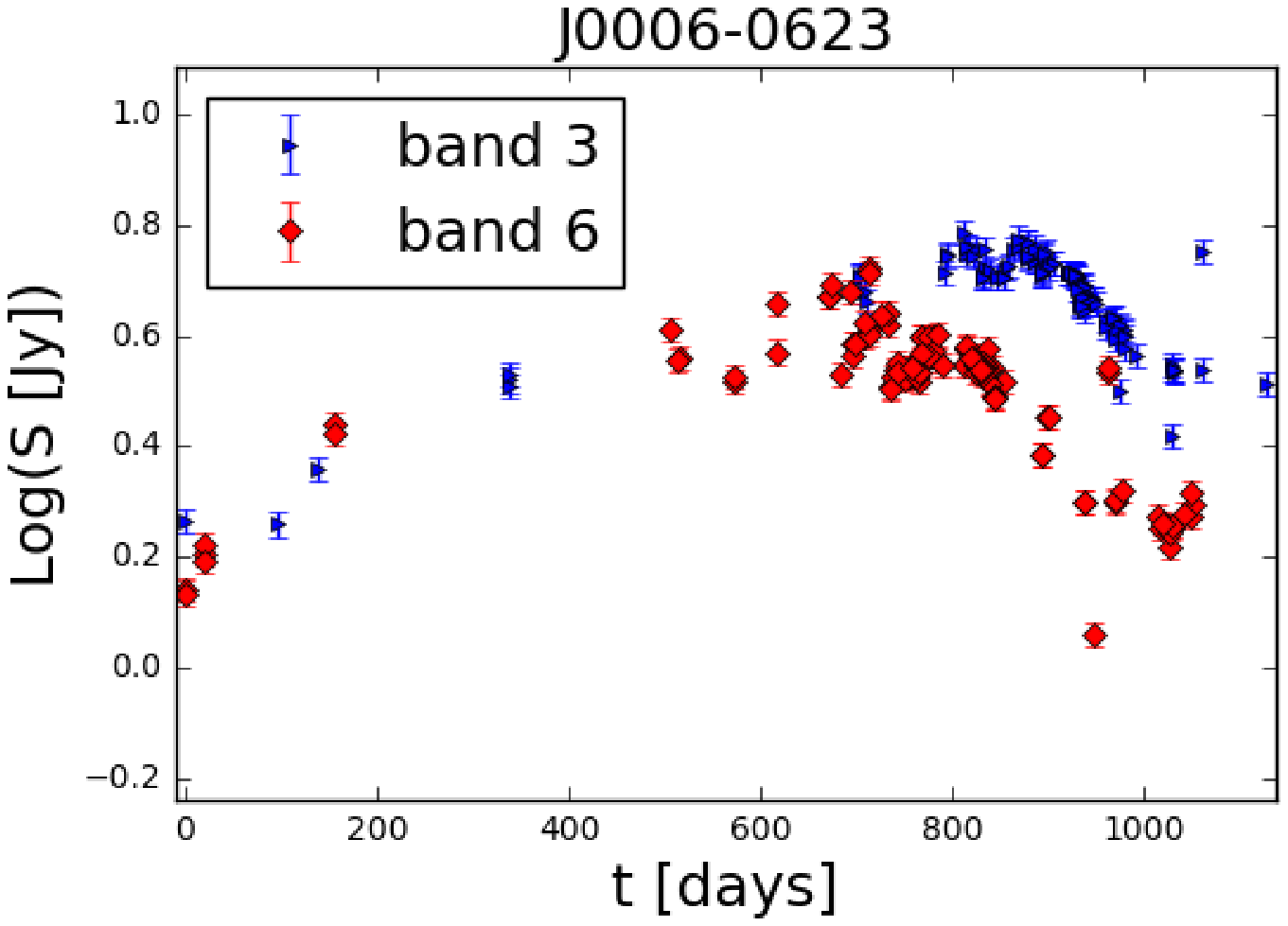}
\includegraphics[trim=1.0cm 0.8cm 5.8cm 12.0cm,natwidth=610,natheight=642,width=0.32\textwidth]{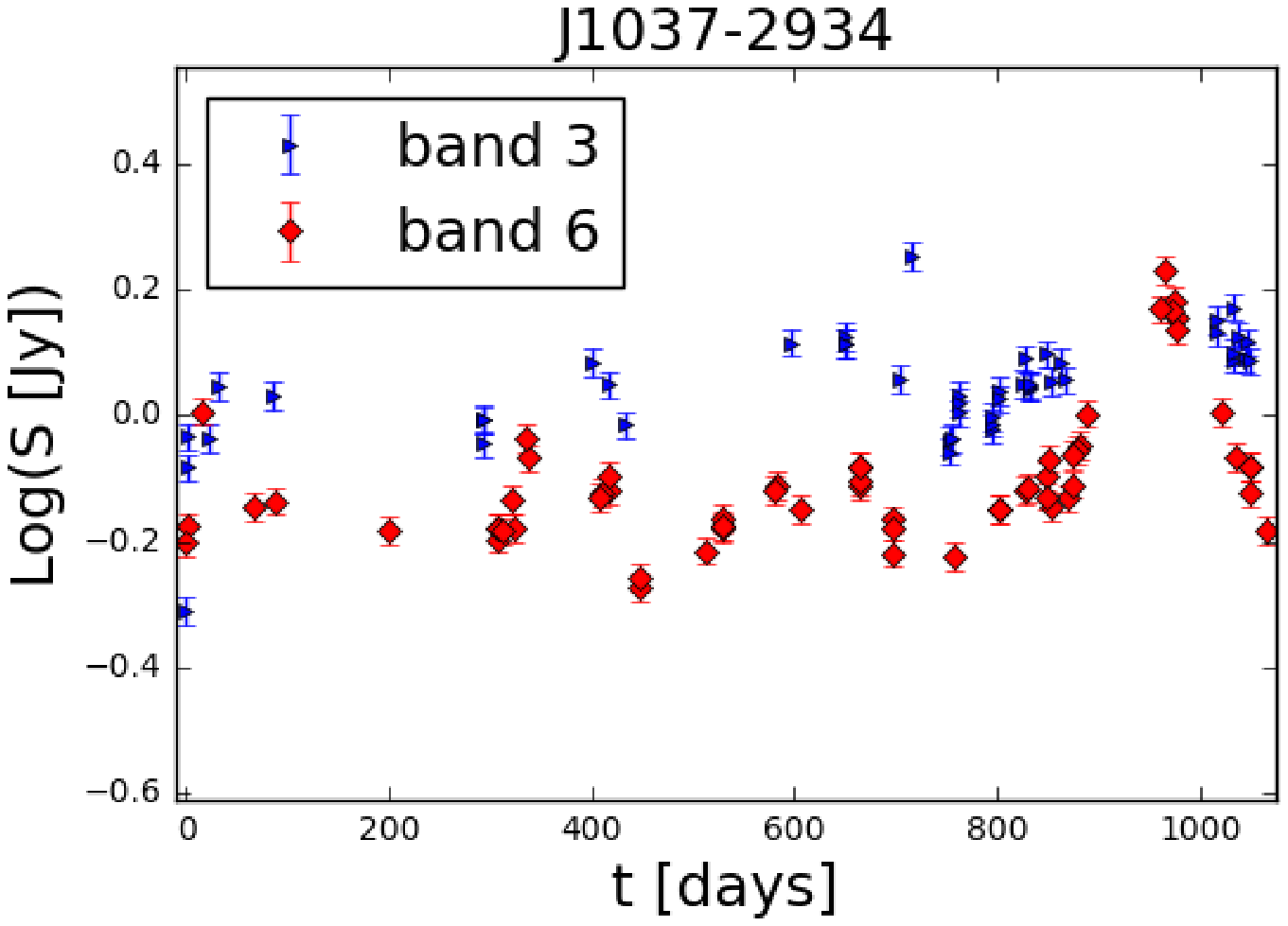}
\includegraphics[trim=1.0cm 0.8cm 5.8cm 12.0cm,natwidth=610,natheight=642,width=0.32\textwidth]{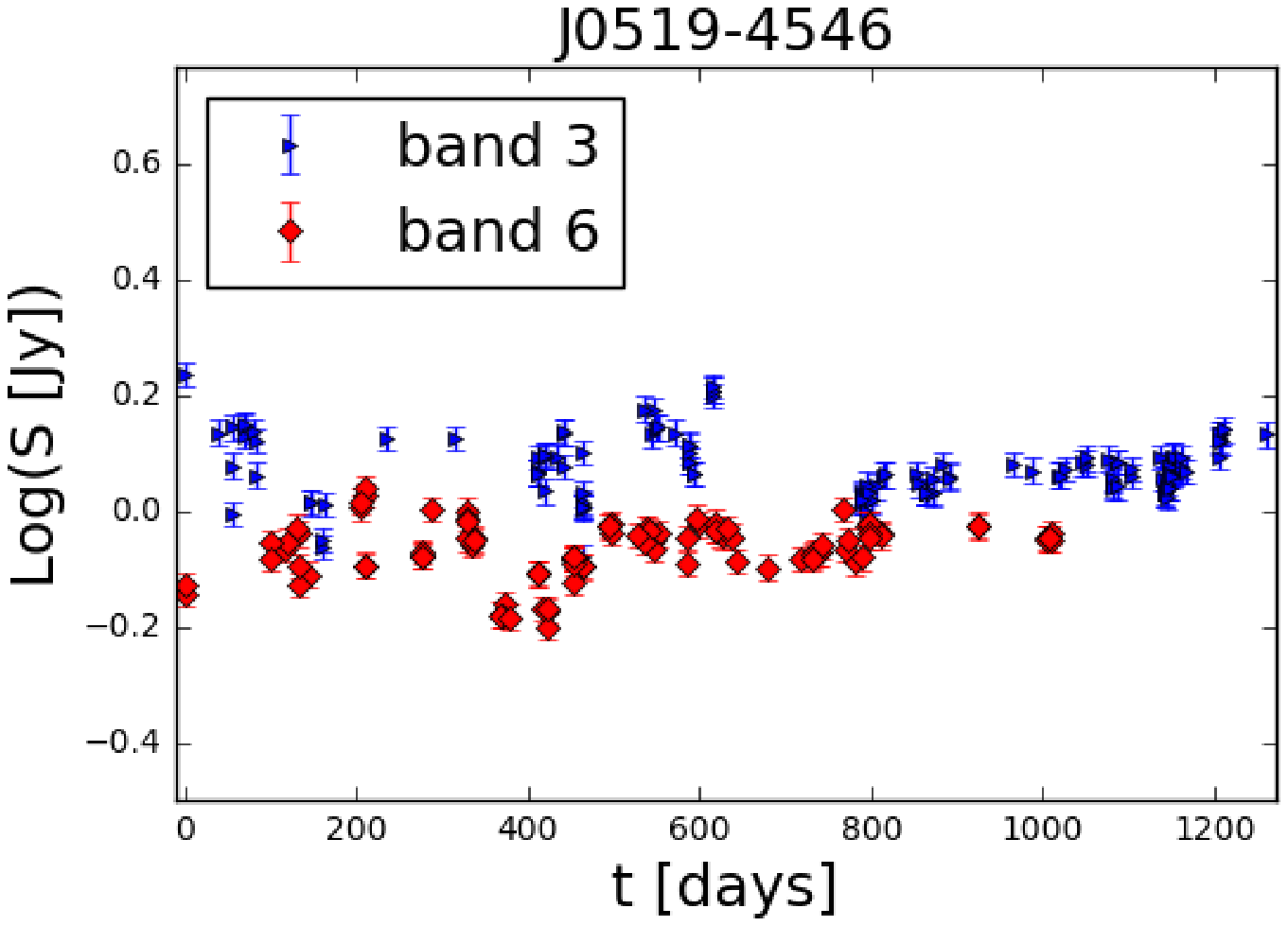}
\caption{Light curves of three of the most frequently observed sources in our sample. The time is in the source frame. All three sources show significant variability but with remarkably different flare morphology.  In the case of J0006$-$0623 ALMA observations have monitored the entire duration of a big flare, lasting for years, showing that it had a similar amplitude in band 6 and in band 3 but a shorter duration in band 6. J1037$-$2934 shows, in both bands, a sequence  of relatively short duration, moderate amplitude flares, the last of which, best monitored  in band 6,  is the most prominent one.  The rise times of the flares are generally shorter than the decay times, consistent with the results by \citet{Nieppola2009}. Also J0519$-$4546 shows a sequence of moderate amplitude, short flares, but they are followed by a relatively quiescent period.}
 \label{fig:LC}
  \end{center}
\end{figure*}


\section{Source classification}\label{sect:sed_classification}

Most sources of our catalogue (489, i.e. $\simeq$67 per cent) are included  in
the 5th edition of the Roma Multi-frequency Catalogue of
Blazars\footnote{\url{http://www.ssdc.asi.it/bzcat/}}
\citep[BZCAT;][]{Massaro2009}. BZCAT sources are divided into 5 sub-classes:
FSRQs, BL Lacs, BL Lacs-galaxy dominated, Blazars of uncertain type, BL Lac
candidates.

We split the remaining 265 sources into the two classical sub-populations of steep-spectrum
and flat-spectrum sources. As usual, such classification is based on the
low-frequency (between $\simeq 1$ and $\simeq 5\,$GHz) spectral index,
$\alpha_{\rm low}$, adopting $\alpha_{\rm low} = -0.5$ ($S_\nu \propto
\nu^\alpha$) as the boundary value. The flat-spectrum population is essentially made by blazars.

We computed $\alpha_{\rm low}$ using the 1.4 GHz flux densities from the NRAO
VLA Sky Survey \citep[NVSS;][]{Condon1998} complemented with those at 843 MHz
from the Sydney University Molonglo Sky Survey \citep[SUMSS;][]{Mauch2003},
combined with those at 4.85\,GHz from the Green Bank 6\,cm
\citep[GB6;][]{Gregory1996} or from the Parkes-MIT-NRAO
\citep[PMN;][]{GriffithWright1993} survey catalogues.

The low frequency spectral index could be computed for all but 13 sources (out
of 265), that were classified as ``uncertain''. Sources with $\alpha_{\rm
low}<-0.5$ were classified as steep-spectrum, provided they did not show clear
variability or $\gamma$-ray emission. Only 10 sources satisfy the criteria for
a steep-spectrum classification.

The overwhelming majority, 731 sources, i.e. $\sim$97 per cent of the sample,
are classified as blazars (since they belong to the BZCAT catalogue or they
fulfill  the criteria presented above). This includes also those with
$\alpha_{\rm low}<-0.5$ but with statistically significant variability and/or
$\gamma$-ray emission (31 sources). We classify as ``Blazar uncertain type''
our blazars without a BZCAT classification.

The classification assigned to each source is given in the second column of
Table\,\ref{tab:fluxes}. The pie chart (Fig.\,\ref{fig:classification_pie})
illustrates the numerical proportions of sources in the different classes.

The classical physical models of blazars predict a steepening of their radio
spectra at millimeter wavelengths \citep{Kellermann1966, Blandford1979}.
Statistical evidence of such steepening has been reported by several authors
\citep{GonzalezNuevo2008, PlanckCollaborationXIII2011,
PlanckCollaborationXV2011, PlanckCollaborationXLV2016}. The ALMA
data allow us to check this prediction on a much larger sample than was
possible before.

\citet{Tucci2011} went one step further. Their most successful physical
evolutionary models of radio sources entail different distributions of break
frequencies (the frequencies where the spectra steepen),  for BL Lacs
and FSRQs. They argue that BL Lacs have substantially higher break frequencies,
implying that their synchrotron emission comes from more compact regions. Their
best model, C2Ex, that successfully fits number counts and spectral index
distributions of extragalactic radio sources over the 5--220\,GHz frequency
range, predicts, for bright blazars ($S_{5\rm GHz}> 0.1\,$Jy, like sources in
our sample), that the break frequencies of most FSRQs are well below 100\,GHz
while those of most BL Lacs are well above this frequency (cf. their Fig.~7).
The ALMA data are well suited to test this prediction.

The left panel of Fig.\,\ref{fig:color_color} shows the distribution of
high-frequency spectral indices ($\alpha_{\rm high}$, from the effective
frequencies of ALMA band 3 [84$-$116\,GHz] to those of band 6 [211$-$275\,GHz])
versus $\alpha_{\rm low}$ for FSRQs and BL Lacs. The right panel shows the
distributions of such $\alpha_{\rm high}$ indices.

Most low-frequency spectral indices are in the range from $-0.5$ to
0.8, while most of the high-frequency ones range from $-1.3$ to 0. Within these
ranges there is no correlation between the high- and low-frequency spectral
indices. The median spectral indices substantially steepen from low to high
frequencies. For FSRQs we have $\alpha_{\rm low, median}\simeq0.11$ (with first
and third quartile values of about $-$0.10 and 0.29) and $\alpha_{\rm high,
median}\simeq-0.65$ (first and third quartile values of about $-$0.85 and
$-$0.51). For BL Lacs $\alpha_{\rm low, median}\simeq0.05$ (first and third
quartile values of about $-$0.18 and 0.22) and $\alpha_{\rm high,
median}\simeq-0.48$ (first and third quartile values of about $-$0.61 and
$-$0.33). The global (FSRQ + BL Lac) median high-frequency spectral index
$\alpha_{\rm high, median}\simeq-0.63$ (with first and third quartile values of
about $-$0.80 and $-$0.45) is in good agreement with those found by
\citet{Massardi2016} for the \textit{Planck}--ATCA Co-eval Observations (PACO)
bright sample: $\alpha_{\rm median, 100-143\,GHz}=-0.67$ (with first and third
quartile values of $-0.94$ and $-0.45$); $\alpha_{\rm median,
143-217\,GHz}=-0.57$ (with first and third quartile values of $-0.83$ and
$-0.45$).

There is thus evidence of a flatter median $\alpha_{\rm high}$ of BL Lacs compared to FSRQs.
The statistical significance of the difference was estimated using the
two-sample Kolmogorov-Smirnov (KS) test, i.e. computing
\begin{equation}
X^2=4\,D^2\,\frac{mn}{m+n}
\label{eq:Siegel}
\end{equation}
where $D$ is the KS statistics, that is the largest discrepancy between the
cumulative distributions of high-frequency spectral indices of the two source
populations, FSRQs and BL Lacs, comprising $m=117$ and $n=36$ sources,
respectively.

We find $D = 0.346$  corresponding to a $0.2$ per cent probability that the two
populations are drawn from the same parent distribution. A simpler, although less rigorous, illustration of the significance of the difference can be obtained considering that the ratio of the numbers of FSRQs in the bins $-1.1\lesssim\alpha_{\rm high}\lesssim-0.5$ and $-0.5\lesssim\alpha_{\rm high}\lesssim0$) is $\sim2.4$ (the FSRQs in the two bins are $71+30=101$; see the right panel of Fig.\,\ref{fig:color_color}). If the $15+20=35$ BL Lacs were extracted from the same parent population we would expect a similar ratio between the two bins, i.e. the expected number of BL Lacs in the first bin would be $71\times(35/101)\sim24.6$ and in the second bin would be $30\times(35/101)\sim10.4$. Based on the Poisson statistics, the probability of getting in the second bin 20 objects when 10.4 are expected is $\simeq 0.3$ per cent, close to the result of the KS test.

The statistically significant difference between the distributions of
$\alpha_{\rm high}$ for the two populations might be consistent with higher
break frequencies for BL Lacs compared to FSRQs, as suggested by the
\citet{Tucci2011} model. However \citet{PlanckCollaborationXLV2016} did not
find significant differences in the break frequencies of the two populations
for their complete flux-density-limited sample of 104 extragalactic radio
sources detected by the \textit{Planck} satellite, but reported average
spectral indices above the break frequency significantly steeper for FSRQs than
for BL Lacs.
\begin{figure*}
\begin{center}
\includegraphics[trim=1.0cm 0.3cm 5.8cm 11.5cm,natwidth=610,natheight=642,width=0.32\textwidth]{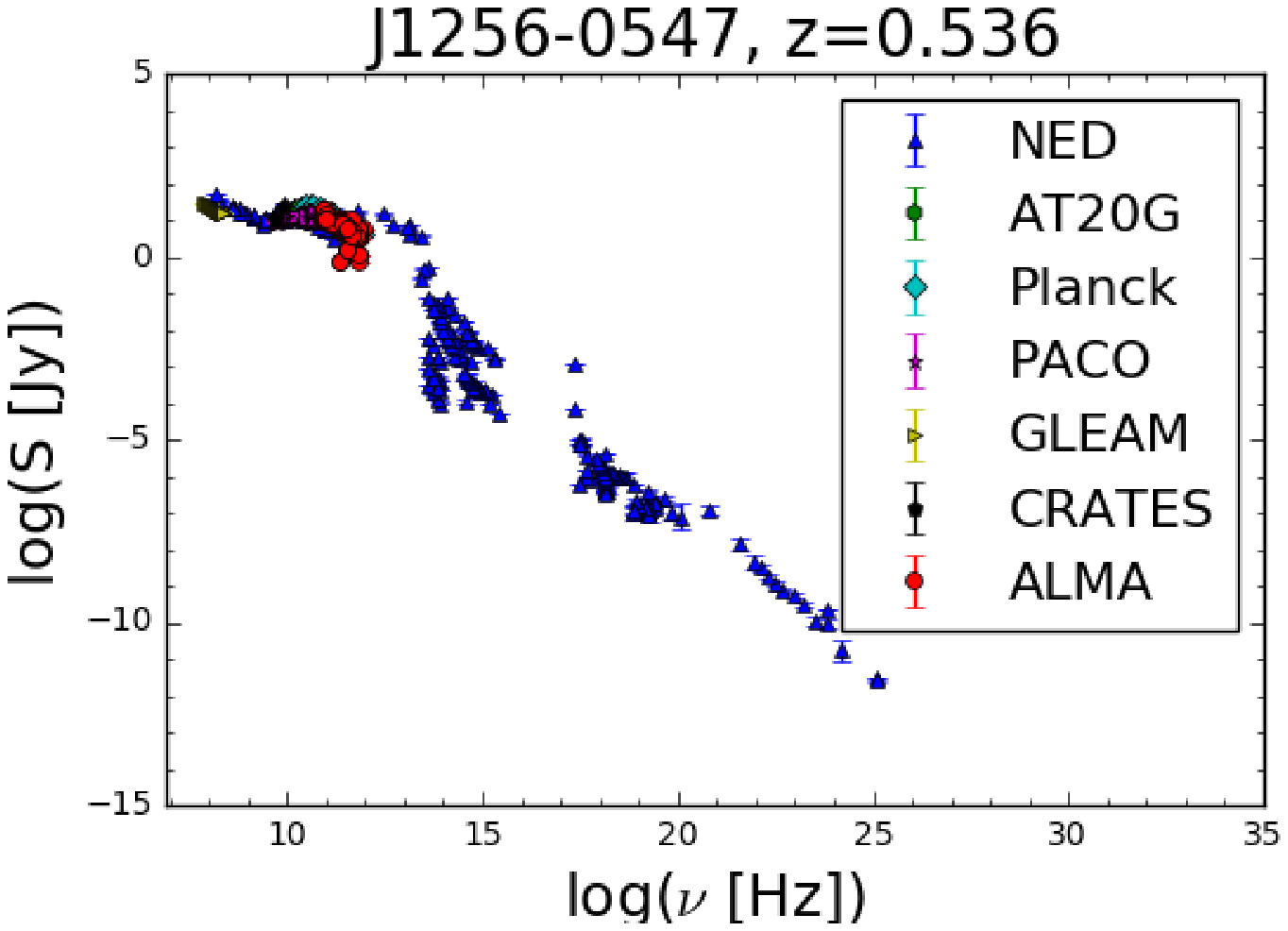}
\includegraphics[trim=1.0cm 0.3cm 5.8cm 11.5cm,natwidth=610,natheight=642,width=0.32\textwidth]{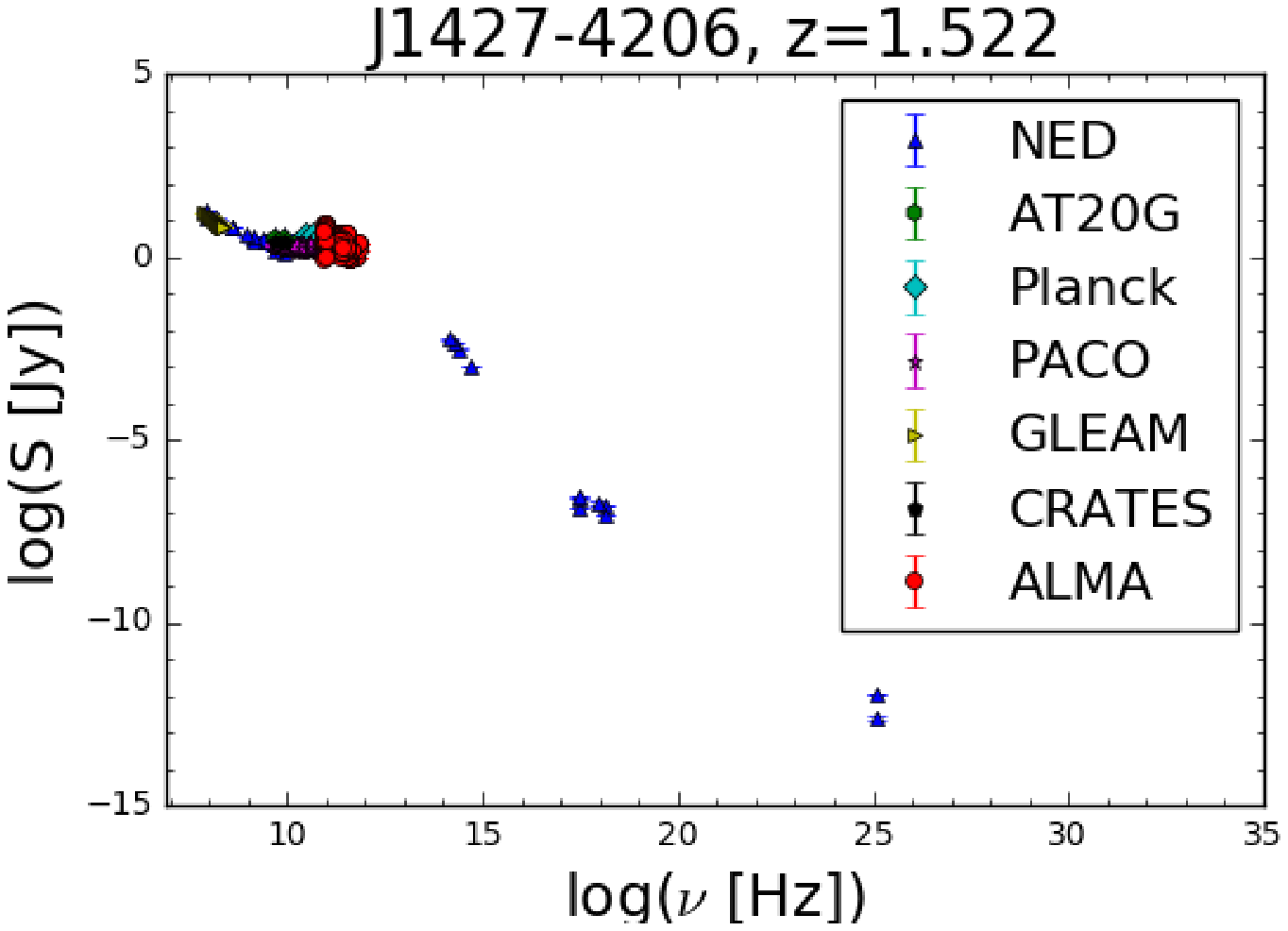}
\includegraphics[trim=1.0cm 0.3cm 5.8cm 11.5cm,natwidth=610,natheight=642,width=0.32\textwidth]{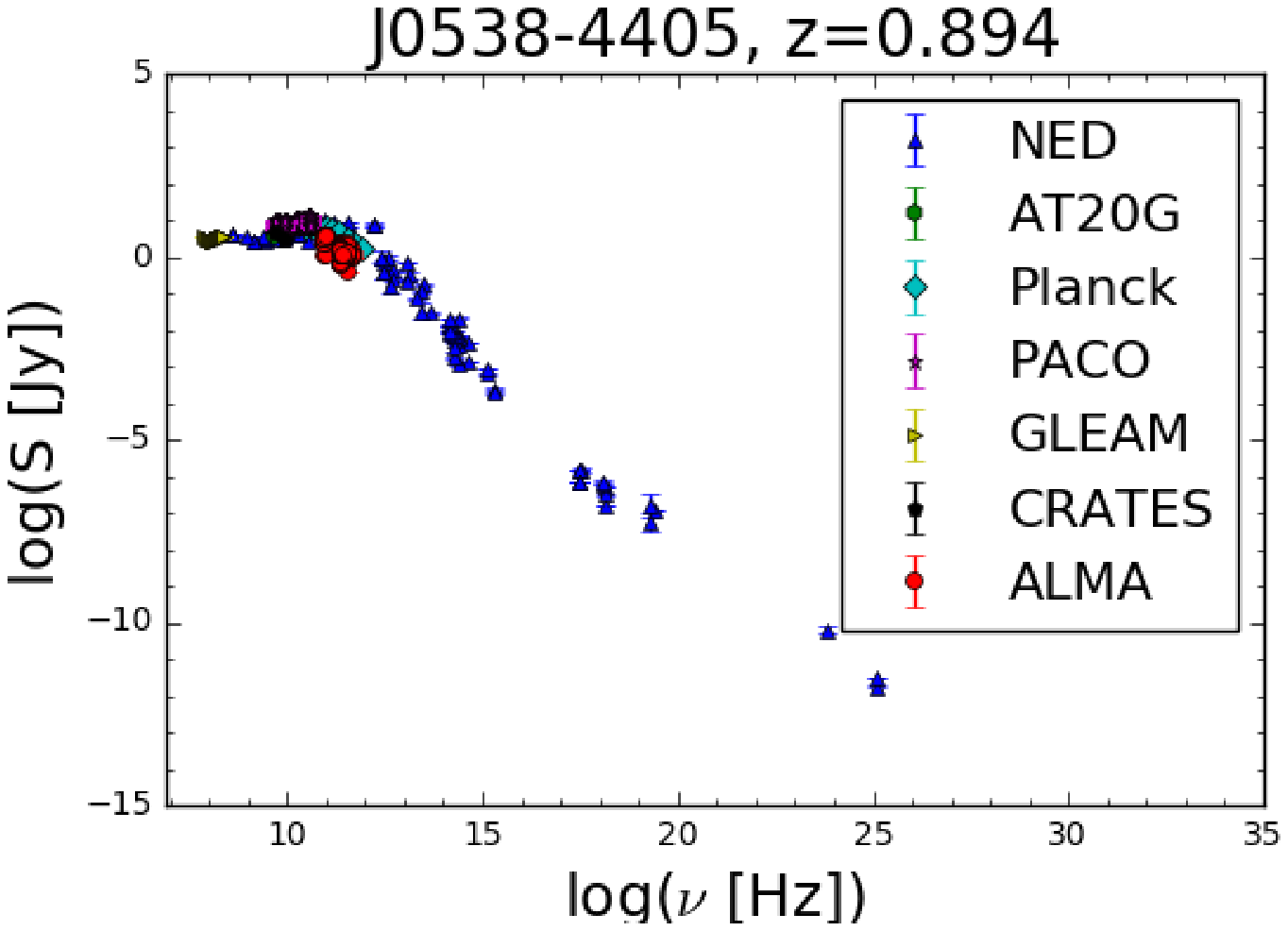}
\includegraphics[trim=1.0cm 0.3cm 5.8cm 11.5cm,natwidth=610,natheight=642,width=0.32\textwidth]{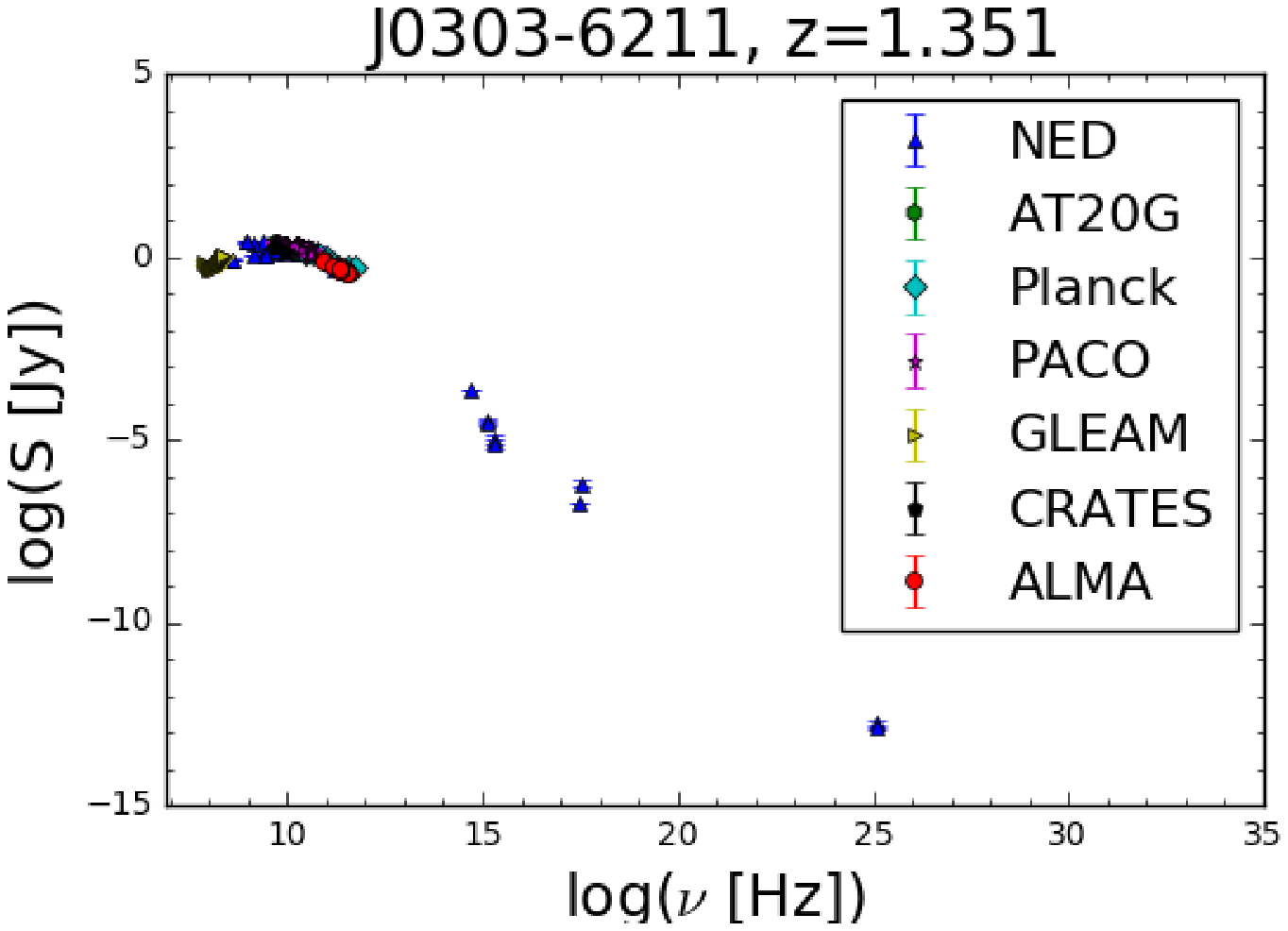}
\includegraphics[trim=1.0cm 0.3cm 5.8cm 11.5cm,natwidth=610,natheight=642,width=0.32\textwidth]{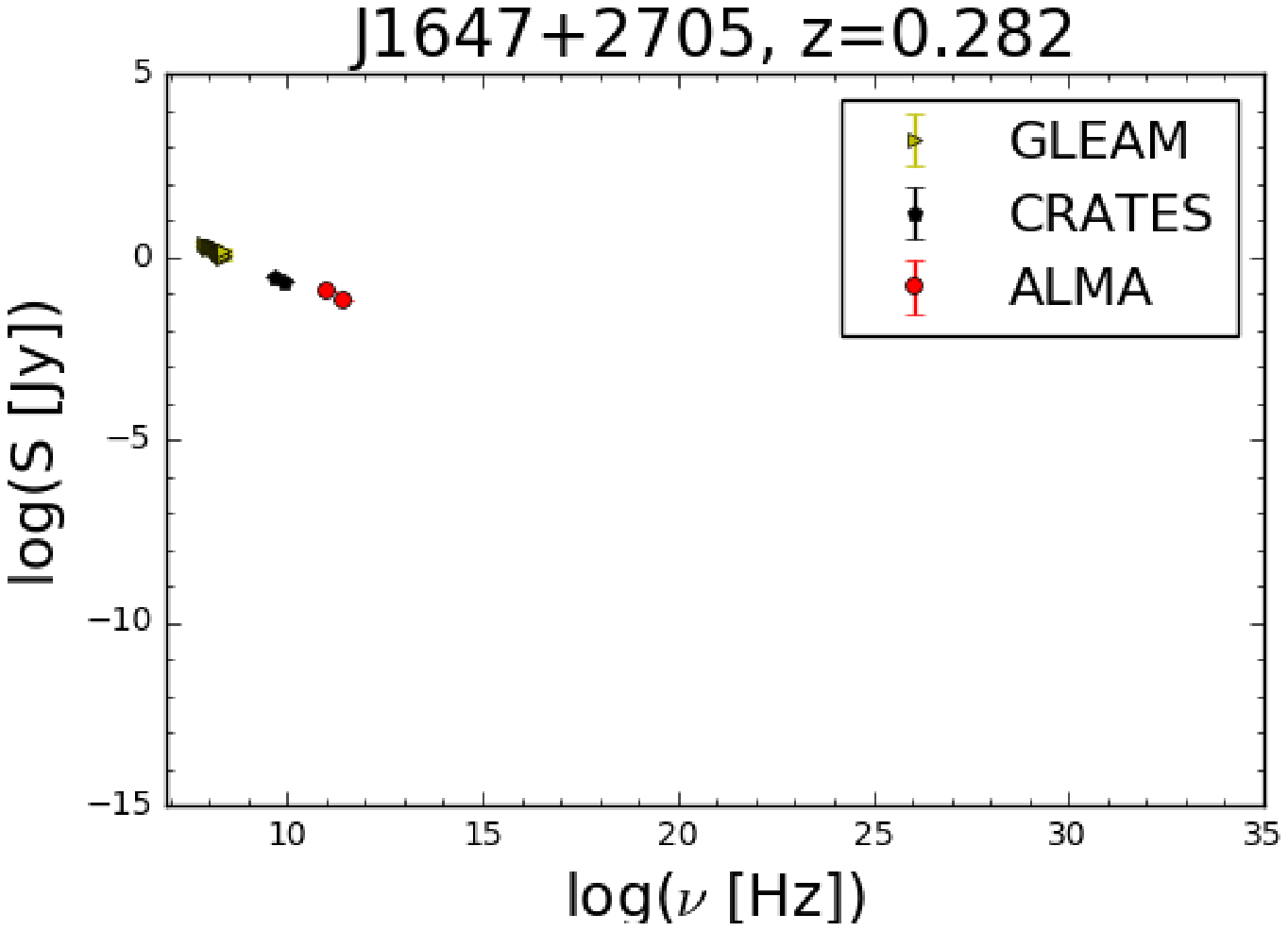}
\includegraphics[trim=1.0cm 0.3cm 5.8cm 11.5cm,natwidth=610,natheight=642,width=0.32\textwidth]{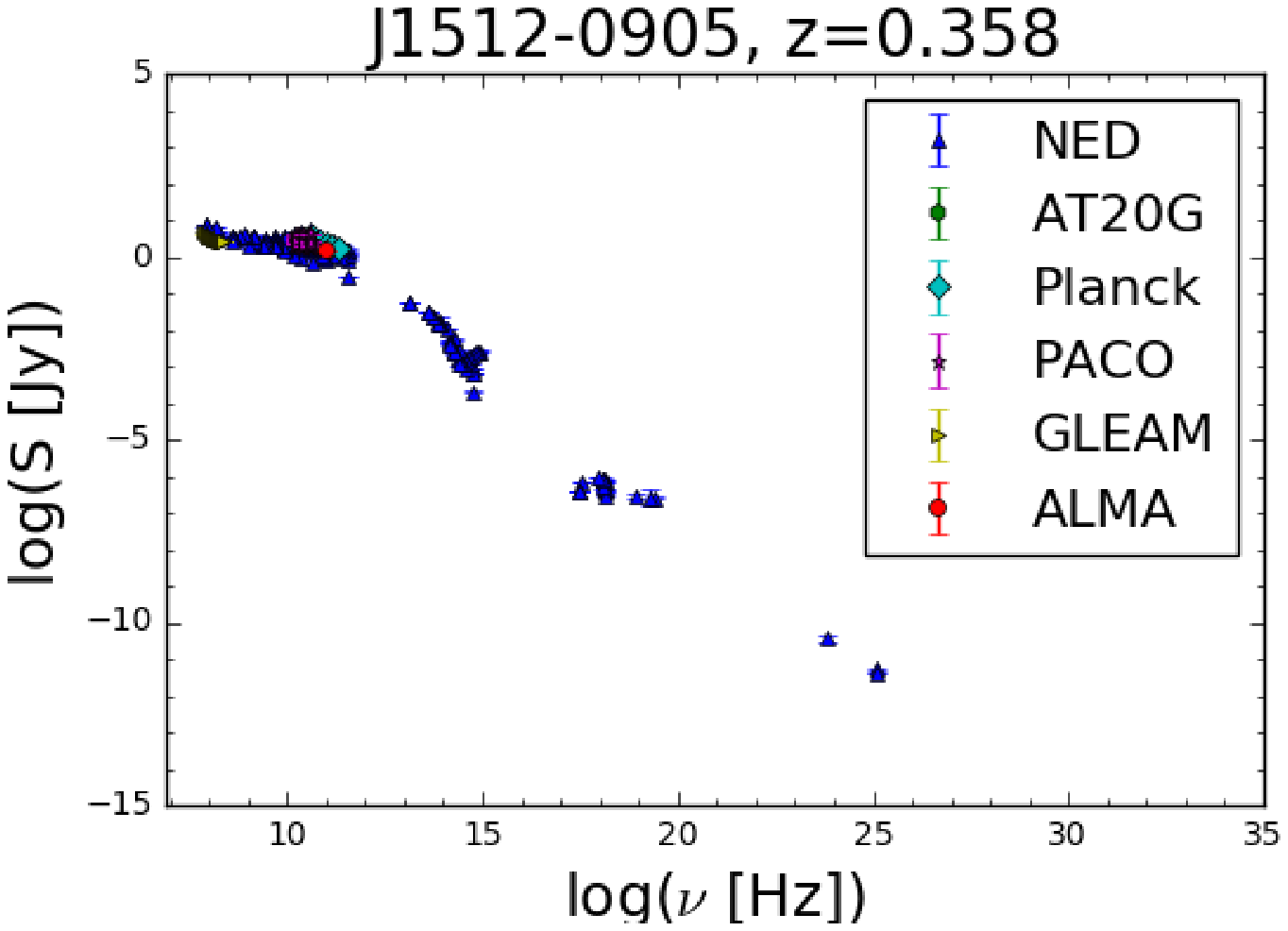}
\caption{Examples of SEDs of our sample reconstructed using the collection of
photometric data described in Sect.\,\ref{sect:sed_extension}: `NED' from the
NASA/IPAC Extragalactic database (\url{https://ned.ipac.caltech.edu/}); `AT20G' from \citet{Murphy2010}; `PCCS2' from \citet{PCCS2}; `PACO' from \citet{Massardi2016}; `GLEAM' from \citet{HW2017}; `CRATES' from \citet{Healey07}; `ALMA' are the new observations presented in this paper. The complete SED collection for our sample is available on the website of the Italian ARC (\url{http://arc.ia2.inaf.it}).}
 \label{fig:SED}
  \end{center}
\end{figure*}

\section{Properties of the sample }\label{sect:properties}

The flux density distributions of sources in the two most frequented ALMA bands
(bands 3 and 6) are shown in Fig.\,\ref{fig:F_distribution}. They extend from
$\sim$1\,mJy to $\sim$15\,Jy, with a peak at $\sim$0.2\,Jy. For comparison, the
minimum flux densities of sources in the ``extragalactic zone'' ($|b|>
30^\circ$) listed in the Second \textit{Planck} Catalogue of Compact Sources
\citep[PCCS2;][]{PCCS2} are 232\,mJy at 100\,GHz and 127\,mJy at 217\,GHz.
Therefore, the ALMA observations reach much fainter flux densities than
the Planck ones, but there is a large overlap between the two sets of
observations.

As mentioned in Sect.\,\ref{sect:data} we have recovered redshift measurements
for $\simeq$78 per cent of our sample (589 sources). The redshift distributions
of sources detected in bands 3 and 6 are shown in
Fig.\,\ref{fig:z_distribution}. Both distributions peak at $0.5<z<1$ and have
tails extending up to $z\sim 3.5$.

Figure\,\ref{fig:N_distribution} shows the distribution of the number of
observations of a given source, again in bands 3 and 6, for sources with
$\geq$2 observations. The distributions  peak at the lowest bin (2-8
measurements) but there is a significant number of sources with tens of
measurements, up to $\sim 250$; sources with $\geq$20 observations are 46 in
band 3 and 52 in band 6.

The time span distribution of measurements in the source frame ($\Delta t_{\rm
source}= \Delta t_{\rm observer}/(1+z)$, where $\Delta t_{\rm observer}$ is the
time between the first and the last observation) is shown in
Fig.\,\ref{fig:time_distribution}.  We excluded all the sources lacking
redshift measurements. Although for most sources the time span is relatively
short (less than a few hundreds of days), for some sources observations cover a
few years (in the source frame); sources with $\Delta t_{\rm
source}\geq2$\,years are 20 in band 3 and 28 in band 6.

The light curves in bands 3 and 6 of three of the most frequently observed
sources are shown in Fig.~\ref{fig:LC}. Monitoring of blazars is important to
understand which mechanisms drive their violent variability and what is the
duty cycle of their activity. The interest on multi-frequency blazar monitoring
has gained momentum since the launch of the \textit{Fermi} Gamma-ray Space
Telescope. In fact the overwhelming majority of detected extragalactic
$\gamma$-ray sources are blazars and \textit{Fermi} has gathered spectacular
$\gamma$-ray light curves of hundreds of them \citep{Abdo2010, Abdollahi2017}.
The poorly explored few-mm to sub-mm spectral region covered by ALMA
observations is important in this context since flux densities in this region
appear to be particularly well correlated with those at 1\,GeV
\citep{Fuhrmann2016}.

An obvious use of multiple observations is the calculation of the variability
index ($VI$), defined as \citep{Sadler2006}:
\begin{equation} \label{eq:VI}
VI=\frac{100}{\langle S \rangle}\times\sqrt{\frac{\sum[S_{i}-\langle S \rangle]^2-\sum(\sigma_{i})^2}{N}}
\end{equation}
where $S_{i}$ and $\sigma_{i}$ are the flux density measurements of a source
measured in a given band and the associated uncertainties, $N$ is the number of
measurements, and $\langle S \rangle$ is the mean flux density.


Obviously, the variability index can be reliably  measured only if the
amplitude of flux density variations is substantially larger than the 5 per
cent calibration uncertainty, although Eq.~(\ref{eq:VI}) gives values of the
variability index $<$5 per cent. Reliable variability indices are measured for
31 out of 41, 30/33, 29/33 and 26/26 (band\,3) and 25/30, 36/39, 35/39 and
37/37 (band\,6) sources for time spans of 100, 200, 400 and 800 days (within
$\pm$30 per cent, in the source frame), respectively. All sources with $\ge 2$
measurements on these timescales are included.

\begin{table}
\centering
\begin{tabular}{ccccc}
\hline
\hline
Band & Time span & VI$_{\textrm{median}}$ & VI$_{1^{st} \textrm{quartile}}$ & VI$_{3^{rd} \textrm{quartile}}$ \\
& [days] & [\%] & [\%] & [\%]\\
\hline
3 & 100 & 5.9  & 1.7  & 11.8 \\
& 200 & 12.8  & 5.7  & 24.3 \\
& 400 & 14.2  & 4.5  & 24.8 \\
& 800 & 23.9  & 13.9  & 32.8 \\
\hline
6 & 100 & 8.1  & 4.4  & 12.2 \\
& 200 & 11.9  & 7.0  & 20.2 \\
& 400 & 14.3  & 8.1  & 25.6 \\
& 800 & 21.3  & 18.6 & 33.4 \\
\hline
\hline
\end{tabular}
\caption{Median, first quartile and third quartile values of the variability indices for 4 different source frame time spans (100, 200, 400 and 800 days) in ALMA band 3 and 6.} \label{tab:VI}
\end{table}

In Table\,\ref{tab:VI} the median, first quartile and third quartile values of
the VI for the different source frame timescales and in the two different bands
are listed.


Applying the two-sample KS test to the  VI distributions of the 100 and 800
days of timescales, we find that the probability that the two sub-samples are
drawn from the same parent distribution is extremely low ($<$0.1 per cent in
both bands, with D$=$0.588 in band 3 and D$=$0.732 in band 6). This is a direct
consequence of the fact that the characteristic timescale of blazar variability
in blazar sources is $\sim$3\,years (see e.g. \citealt{Nieppola2009}).

Measured flux densities of the same source for short time spans are
expected to be only weakly affected by variability. Differences among such
measurements are therefore an estimator of systematic errors that afflict our
observations, and primarily of the calibration error. The median absolute values
of differences among measurements in bands 3 and 6 done within 30 days in the
source frame are of $\simeq 4$ per cent for band 3 and of $\simeq 5$ per cent
for band 6, consistent with the adopted calibration error of 5 per cent.


\section{Global spectral energy distributions (SEDs) of sources}\label{sect:sed_extension}

As mentioned in Sect.~\ref{sect:sed_classification}, our source classification
is based on external data. For each source in our sample, we have collected the
photometric data available on the NASA/IPAC Extragalactic Database (NED) using
Astroquery with a search radius of 10 arcsec and excluding Galactic sources.

The NED data were complemented by cross-matching our  catalogue with: the
Australia Telescope 20GHz Survey Catalog (AT20G; \citealt{Murphy2010}), the
PCCS2 (\citealt{PCCS2}), the PACO catalogue (\citealt{Massardi2016}), the
GaLactic and Extragalactic All-sky MWA survey (GLEAM; \citealt{HW2017}), the
CRATES survey (\citealt{Healey07}).

The cross-matching was done using the following search radii: 5 arcsec around
the ALMA positions for the AT20G catalog; 16, 13.5, 6.5, 4.85, 3.6, 2.45, 2.45,
2.35 and 2.1 arcmin (i.e. half FWHM) for the 30, 44, 70, 100, 143, 217, 353,
545 and 857\,GHz PCCS2 catalogues, respectively; 20 arcsec for the GLEAM
catalog; 70 arcsec for the CRATES one. For the PACO catalog we exploited the
AT20G identifications by \citet{Massardi2016}. We considered PCCS2 data only
for sources with $|b|>10^\circ$ to avoid wrong identifications (with Galactic
sources). In all the cases, the search yielded a unique identification.

In this way we obtained SEDs extending over 17 orders of magnitude, from radio
to $\gamma$-rays. The complete SED collection is  available on the website of
the Italian ARC (\url{http://arc.ia2.inaf.it}). Some examples are shown in
Fig.\,\ref{fig:SED}. We found $\gamma$-rays measurements for 248 sources
($\sim$33 per cent of the sample).

\section{Summary}\label{sect:conclusions}

We have presented a catalogue of ALMA flux density measurements of calibrators,
observed between August 2012 and September 2017, in the framework of the ALMACAL project. The ALMACAL images were
reprocessed using a new code developed by the Italian node of the European ALMA
Regional Centre. This has yielded 16,263 flux density measurements in different
ALMA bands and at different epochs of 754 calibrators. A search in online
databases has yielded redshifts for 589 sources ($\sim$78 per cent of the total).

Most (489, i.e. $\simeq$67 per cent) of our sources are classified as blazars
of various types in the BZCAT catalogue. Almost all of the remaining sources
have  properties (flat low-frequency radio spectrum, clear variability in
different bands, $\gamma$-ray emission) consistent with a blazar
classification. In total, $\sim$97 per cent of the sources are classified as
blazars.

To illustrate the properties of the sample, in view of its exploitation for
scientific investigations, we have focussed on the most frequented ALMA bands,
i.e. bands 3 and 6. For these bands we have shown the redshift and flux density
distributions of catalogued sources, the distribution of the number of
observations of individual sources and of time spans in the source frame.

Several sources have tens of measurements in a band, covering several years. As
an example of the variety of scientific investigations allowed by the catalogue,
we have presented unprecedented band 3 and 6 light curves of three sources and
estimates of the variability indices on timescales of 100, 200, 400 and 800
days in the same bands.

Through an analysis of flux density differences for short time spans, in bands 3 and 6, we have found that the systematic errors are consistent with the adopted calibration error of 5 per cent.

We have also found that the ALMA data show highly significant evidence of a
difference between the high-frequency ($\nu \simgt 100\,$GHz) spectra of FSRQs
and BL Lacs: at wavelengths a few mm the average spectra of BL Lacs are flatter
than those of FSRQs. This is expected if the synchrotron emission of BL
Lacs comes from more compact regions than the emission of FSRQs, as argued,
e.g., by \citet{Tucci2011}.

Finally, by collecting data from online databases, we have reconstructed the
SEDs of our sources over 17 orders of magnitude in frequency. Both the
catalogue and the SEDs are available to the community.

\section*{Acknowledgements}
We thank Ian Smail for his useful comments. We are indebted to Edward
Fomalont for enlightening discussions, highly valuable suggestions and for his
extensive work, together with Ruediger Kneissl and Antonio Hales, on ALMA
calibrator catalogs and calibration error estimation. This paper and the AKF
and KAFE development are part of the activities for the ALMA Re-Imaging Study
approved in the framework of the 2016 ESO Call for Development Studies for ALMA
Upgrade (PI: Massardi). The study acknowledges partial financial support by the
Italian Ministero dell'Istruzione, Universit\`{a} e  Ricerca through  the grant
`Progetti Premiali 2012 - iALMA' (CUP C52I13000140001). GDZ acknowledges
support from ASI/INAF agreement n.~2014-024-R.1 for the {\it Planck} LFI
Activity of Phase E2 and from the ASI/Physics Department of the university of
Roma--Tor Vergata agreement n. 2016-24-H.0 for study activities of the Italian
cosmology community. MN acknowledges support from the European Union's Horizon
2020 research and innovation programme under the Marie Sk{\l}odowska-Curie
grant agreement No 707601.  This research has made use of the NASA/IPAC
Extragalactic Database (NED), which is operated by the Jet Propulsion
Laboratory, California Institute of Technology, under contract with the
National Aeronautics and Space Administration.

\bibliographystyle{mnras}
\bibliography{biblio} 

\bsp	
\label{lastpage}
\end{document}